\documentclass[useAMS,usenatbib]{mn2e}

\usepackage{graphicx}

\title[The colliding winds of $\eta$ Carinae]
{Wind-wind collision in the $\eta$ Carinae binary system: a shell-like event near %%@
periastron}
\author[D. Falceta-Gon\c{c}alves, V. Jatenco-Pereira and Z. Abraham]{D. 
Falceta-Gon\c{c}alves$^{1}$\thanks{E-mail:
diego@astro.iag.usp.br}, V. Jatenco-Pereira$^{1}$ and Z. Abraham$^{1}$\\
$^{1}$Instituto de Astronomia, Geof\'\i sica e Ci\^encias Atmosf\'ericas, %%@
Universidade de S\~ao Paulo, Rua do Mat\~ao 1226, CEP 05508-900, S\~ao Paulo, Brazil}
\begin{document}

\date{}

\pagerange{\pageref{firstpage}--\pageref{lastpage}} \pubyear{2004}

\maketitle

\label{firstpage}

\begin{abstract}
The exact nature of $\eta$ Carinae is still an open issue. Strict periodicity in the %%@
light curves at several wavelengths seem to point out to a binary system, but the %%@
observed radial velocities, measured from space with high spatial resolution are in %%@
conflict with the ground based observations used to calculate the binary orbit. %%@
Also, the observed 2-10 keV X-ray flux is much larger that what is expected from a %%@
single star, and favors the wind-wind collision hypothesis, characteristic of high %%@
mass binary systems. However, to explain the duration of the dip in the light curve %%@
by wind collisions, it is necessary to postulate a very large increase in the $\eta$ %%@
Carinae mass loss rate. Finally,  the optical and UV light curves are better %%@
explained by periodic shell-ejection events. In this paper we conciliate the two %%@
hypothesis. We still assume a binary system to explain the strong X-ray emission, %%@
but we also take into account that, near periastron and because of the highly %%@
eccentric orbit, the wind emerging from $\eta$ Carinae accumulates behind the shock %%@
and can mimic a shell-like ejection event. For this process to be effective, at %%@
periastron the secondary star should be located between $\eta$ Carinae and the %%@
observer, solving also the discrepancy between the orbital parameters derived from %%@
ground and space based observations. We show that, as the secondary moves in its %%@
orbit, the shell cools down and the number of available stellar ionizing photons is %%@
not enough to maintain the shell temperature at its equilibrium value of about 7500 %%@
K. The central part of the shell remains cold and under these conditions grain %%@
formation and growth can take place in timescales of hours. We also calculated the %%@
neutral gas column density intercepting the line of sight at each point of the orbit %%@
near periastron, and were able to reproduce the form and duration of the X-ray light %%@
curve without any change in the $\eta$ Carinae mass loss rate. This same column %%@
density can explain the observed H$\alpha$ light curve observed during the 2003 %%@
event.
\end{abstract}


\section{Introduction}

$\eta$ Carinae is a peculiar star surrounded by a dense cloud of gas and dust, %%@
formed in several episodes of mass ejection, typical of  LBV stars. Its mass, %%@
calculated from the bolometric luminosity, corresponds to approximately 120 %%@
M$_{\odot}$, making $\eta$ Carinae, if single, the most massive star in the Galaxy. 
During the last years strong evidences of the system binary nature  were reported: a %%@
5.52 year periodicity in the light curve at infrared wavelengths %%@
\citep{whitelock94}, in the HeI $\lambda$10830 \AA$\;$  high excitation line %%@
\citep{dam96} and in the radial velocity of the Pa$\gamma$ line, the latter pointing %%@
out to a highly eccentric orbit \citep*{damineli97, davidson97}. 
The 5.52-year periodicity was also found at centimeter and millimeter wavelengths %%@
\citep{coxa95,abra99,duncan03,abr03}. 
VLBI observations showed a very compact emitting region, coincident with the star %%@
position at the minimum in the light curve, which grew to an extended and elongated %%@
feature at maximum \citep*{duncan97}, indicating the probable existence of an %%@
ionized disk of variable size. 
Infrared observations also showed a compact disk, with a power law spectrum between %%@
2 and 8 $\mu$m, attributed to thermal emission by dust at temperatures between 1500 %%@
and 400 K \citep{morris99}.

A draw back in the binary system hypothesis came from the high resolution $HST$ %%@
spectroscopic observations of $\eta$ Carinae \citep{davidson00}, which showed that, %%@
although the radial velocity of the emission lines formed close to the star varied %%@
at the different phases of the binary orbit, they were incompatible with the orbital %%@
parameters derived by \citet{damineli97}. 
Without the assumption of a binary companion, the spectroscopic events could be %%@
explained if they were the result of periodic shell ejections %%@
\citep*{zanella84,davidson99, martin04}. 
Under these conditions, the dense gas in the shell would screen the stellar UV %%@
radiation, responsible for the formation of the ionized envelope surrounding $\eta$ %%@
Carinae, favoring also the formation of dust, which could explain the enhanced IR %%@
emission.  

On the other hand, a strong argument in favor of the binary hypothesis is based in  %%@
the intensity of the observed X-ray emission, which cannot be explained by shell %%@
ejection but can be easily accounted for by free-free emission from wind-wind %%@
collisions \citep{pittard98}. In fact,  this process was already  postulated in %%@
other massive binary systems to explain the differences between the observed X-ray %%@
emission ($\sim 10^{35}$ erg s$^{-1}$) and what is expected from individual stars %%@
($\sim 10^{32}-10^{33}$ erg s$^{-1}$) \citep{prilutskii76, pollock87, zhekov00}.

The 1997 and 2003 events, characterized by  minima in the light curves at other %%@
wavelengths, were also observed at X-rays, preceded by a strong increase in the flux %%@
density. 
The enhanced  emission was  attributed to an increase in the wind-wind collision %%@
strength as the stars approach each other during  periastron passage; the sharp %%@
decrease towards the minima was assumed to be produced by an increase in the H %%@
column density, perhaps as the result of an increase in the $\eta$ Carinae mass loss %%@
rate  \citep{pittard98, corcoran01, pittard02}. 

In this paper we conciliate the two hypothesis: a binary system and a shell %%@
ejection-like event. 
We use the binary system to explain the X-ray emission by shock formation in the %%@
wind-wind collision. 
What we call the shell ejection-like event is the appearence of  dense material %%@
between the observer and the binary system, formed by the stellar wind matter %%@
accumulated behind the cool expanding  shock remnant.  
For this process to be effective in absorbing the X-rays, the periastron must be in %%@
the oposite direction of what was proposed by \citet{damineli00} and %%@
\citet{corcoran01}, being compatible with the $HST$ observations reported by %%@
\citet{davidson00}. 
We show that in this event dust can be formed in the shock interface, where the gas %%@
cools down by radiation and it is shielded from the UV stellar radiation by the %%@
large amount of accumulated gas. 
The dense gas and dust shell will eventually expand and form a disk in the plane of %%@
the orbit, its internal parts will be ionized by the UV radiation of the two stars %%@
in the binary system and will be seen at radio wavelengths as free-free emission. 
The disk will also slow down the wind of $\eta$ Carinae in the plane of the orbit, %%@
which could contribute with other effects (e.g. non symmetrical wind from the %%@
massive rotating primary star) to the  latitude-dependent behavior observed by %%@
\citet{smith03}.

In Section 2 of this paper we calculate the physical conditions in the shocked %%@
region, taking into account radiative absorption and emission. 
In Section 3 we calculate the grain formation rate and in Section 4 we compare our %%@
results with the optical and UV observations  of the 2003 low excitation event %%@
reported by  \citet{martin04}. Finnaly  in Section 5 we present the derived X-ray %%@
light curves for different values of the orbital parameters and of the $\eta$ %%@
Carinae mass loss rate and in Section 6 we present our conclusions. 

\section{The Shock Model}

In massive binary stellar systems, in which both stars present high mass loss rates %%@
in the form of supersonic winds, a shocked region is formed between the two stars, %%@
resulting in an increase of gas density and temperature. 
One of the main observational signatures of this process is the strong X-ray %%@
emission, since in supersonic shocks the gas temperature increases from $10^4$ K to %%@
$\sim 10^7 - 10^8$ K and free-free emission becomes very important;  this kind of %%@
process is now commonly associated to wind collisions in many objects %%@
\citep{corcoran01,thaller01,lepine01}.
The strength of the shock depends on the separation between the stars and it can %%@
increase dramatically  during periastron passage in highly eccentric orbits, as %%@
seems to occur in the $\eta$ Carinae system.

For the study of the shock region we used, as in Chlebowski (1989) and Usov (1992), %%@
the basic hydrodynamics equations for mass continuity, fluid momentum and energy %%@
conservation, neglecting magnetic fields. Considering an adiabatic stationary shock, %%@
the jump conditions are given by:
\begin{equation}
\rho_{1}v_{1}=\rho_{2}v_{2},
\end{equation}
\begin{equation}
P_{1}+\rho_{1}v_{1}^{2}=P_{2}+\rho_{2}v_{2}^{2},
\end{equation}
\begin{equation}
\frac{\gamma}{\gamma-1} %%@
\frac{P_{1}}{\rho_{1}}+\frac{1}{2}v_{1}^{2}=\frac{\gamma}{\gamma-1} %%@
\frac{P_{2}}{\rho_2}+\frac{1}{2}v_{2}^{2},
\end{equation}
where $\rho$ is the gas density, $v$ the relative wind velocity, $P$ the gas %%@
pressure and $\gamma$ the adiabatic exponent; indices 1 and 2 indicate pre and %%@
post-shock parameters. From equations (1) - (3) we can calculate the jump relations %%@
for density and temperature:
\begin{equation}
\frac {\rho_{2}}{\rho_{1}}=\frac{(\gamma +1)M^2}{(\gamma -1)M^2+2},
\end{equation}
\begin{equation}
\frac{\rho_{2}T_{2}}{\rho_{1}T_{1}}=\frac{2\gamma M^2-(\gamma -1)}{\gamma +1},
\end{equation}
where  $M \equiv v_{1}\left( \gamma kT_{1}/\mu m_{H} \right)^{-\frac{1}{2}}$ is the %%@
Mach number, $\mu$ the molecular weight and $m_H$ the mass of the H atom.

 The shock front is formed at a
distance $r_1$ from the primary star given by: 
\begin{equation}
\medskip
r_{1}=\frac{D}{1+\eta^{\frac{1}{2}}}
\medskip
\end{equation}
where $D$ is the distance between the two stars and $\eta=\dot{M_s} v_s/\dot{M_p} %%@
v_p$. $\dot{M_p}$ and $\dot{M_s}$ are the mass loss rates of $\eta$ Carinae and the %%@
companion star, respectively, and $v_p$ and $v_s$ their assimptotic wind velocities.

For the $\eta$ Carinae system, no direct information is available about the %%@
individual masses, compositions and mass-loss rates since, even with the very high %%@
spatial resolution achieved by the {\it HST}, it was not possible to separate the %%@
stellar spectra from that of the surrounding nebula. 
\citet{hillier01}, analyzing the P Cygni velocity profiles, favored a mass-loss rate %%@
of $10^{-3}  \, \rm {M}_{\odot}\; \rm {yr}^{-1}$, while \citet{pittard02} found that %%@
a mass-loss rate of $2.5 \times 10^{-4}\, \rm {M}_{\odot}\; \rm {yr}^{-1}$ fitted %%@
better the X-ray spectra obtained with the {\it Chandra} grating spectrometer.
Observations indicate  that $\eta$ Carinae has a terminal wind velocity of $v_p \sim %%@
500 - 700$ km s$^{-1}$ \citep{hillier01}; its companion is probably an OB or WR type %%@
star \citep{damineli00}, with typical wind parameters, {\it e.g.} $\dot{M_s} \simeq %%@
10^{-5}\; \rm {M}_{\odot}\; \rm {yr}^{-1}$ and $v_s \simeq 3 \times 10^{3}$ km %%@
s$^{-1}$ \citep{abbott86,lamers01,pittard02}.

The orbital parameters are also uncertain. While \citet{damineli97} obtained an %%@
eccentricity $e =0.63$ from the radial velocity of the Pa$\gamma$ and Pa$\delta$ %%@
lines, \citet{davidson97},  analyzing the same data found that an eccentricity of %%@
0.8 fitted better the radial velocity curve. An even higher eccentricity was %%@
postulated by \citet{corcoran01} ($e=0.9$) to reproduce the X-ray light curve. 

Therefore, to calculate the physical parameters at the shock region, we will assume %%@
$e = 0.9$,  $\dot{M_p} = 2.5 \times 10^{-4}\, \rm {M}_{\odot}\; \rm {yr}^{-1}$, %%@
$\dot{M_s} = 10^{-5}\, \rm {M}_{\odot}\; \rm {yr}^{-1}$, $v_p = 700$ km s$^{-1}$ and %%@
$v_s =3000$ km s$^{-1}$.  
Under these conditions, we find that at  periastron $r_1\sim 1.1$ A.U.,  $n_{1p}\sim  %%@
5\times 10^{10}$ cm$^{-3}$ for the $\eta$ Carinae wind and $n_{1s}\sim  1.4\times %%@
10^{9}$ cm$^{-3}$ for the wind of the secondary star, where $n_1=\rho _1/\mu m_H$ is %%@
the gas number density. From equations (4) and (5) and assuming $T_{1}\sim 10^{4}$ K %%@
for both winds, the post-shock density and temperature for $\eta$ Carinae and the %%@
companion star winds turned out to be $n_{2p}\sim 2 \times 10^{11}$ cm$^{-3}$, %%@
$n_{2s}\sim 5.6 \times 10^{9}$ cm$^{-3}$, $T_{2p}\sim 6 \times 10^{6}$ K and %%@
$T_{2s}\sim 2 \times 10^{8}$ K.  These results are compatible with the limits %%@
imposed to the density and temperature by the recent $Chandra$ X-ray spectral %%@
observations \citep{seward01, corcoran01b}.

As a consequence of the high temperature and density at the shock region, the gas %%@
emits  large amounts of radiation and cools down. This process is totally described %%@
by the first law of thermodynamics; considering also isobaric transformations for %%@
gas equilibrium we can write:

\begin{equation}
\frac{3k\rho(t)}{2\mu m_{H}} \frac{dT(t)}{dt}-\frac{kT(t)}{\mu %%@
m_{H}}\frac{d\rho(t)}{dt}+
\vec\nabla\times(K\vec\nabla T)=\Gamma-\Lambda,
\end{equation}

\noindent
where $K$ represents the thermal conduction, $\Gamma$ and $\Lambda$ the energy gain %%@
and loss functions, respectively.  

At high temperatures, the cooling function is dominated by free-free radiation and  %%@
can be expresed by \citep{spitzer78}:  
\begin{equation}
\Lambda_{\rm ff}=2.4\times 10^{-27}T^{\frac{1}{2}}n_i n_e.
\end{equation}

\noindent
where $n_i$ and $n_e$ represent the number density of ions and electrons, %%@
respectively.

At lower temperatures line transitions become important; the cooling function due to %%@
a  transition $j\rightarrow k$ of energy $E_{jk}$, for an element with 
abundance $a_i$ is:
\begin{equation}
\frac{\Lambda_{\rm %%@
line}}{n_{e}n_{p}}=a_{i}x_{j}E_{jk}\gamma_{jk}\left(1+\frac{n_{e}\gamma_{kj}}
{A_{kj}}\right)^{-1},
\end{equation}
where $x_{j}$ is the element  degree of ionization, $\gamma_{jk}$ and $\gamma_{kj}$ %%@
are the collisional excitation and de-excitation parameters and $A_{kj}$ is the %%@
spontaneous emission rate. An approximate fit for the line cooling function was %%@
given by Matthews \& Doane (1990):

\begin{equation}
\Lambda_{\rm line} \sim \frac{b_1 T^{p}}{(1+b_2T^{q})},
\end{equation}
where $b_1=1.53 \times 10^{-27}$, $b_2=1.25 \times 10^{-9}$, $p=1.2$ and $q=1.85$.

The photoionization heating function (erg s$^{-1}$ cm$^{3}$) is given by %%@
\citep{osterbrock74}:

\begin{equation}
\Gamma_{\rm ph}=n_{\rm {H_{0}}}\int^{\infty}_{\nu_{0}}{\frac{4\pi %%@
J_{\nu}}{h\nu}h(\nu - \nu_{0})\sigma_{\nu}
(\rm{H_{0})d\nu}},
\end{equation}

\noindent
where $\sigma_{\nu} \sim 7.9 \times 10^{-18} Z^{-2} ({\nu_{0}}/{\nu})^{3}g$ is the %%@
photoionization cross section, $Z$ is the
atomic number, $g$ is the Gaunt factor, $\nu_{0}$ is the Lyman limit frequency and %%@
$J_{\nu}$ is the photon mean 
intensity.

\begin{figure}
      {\includegraphics{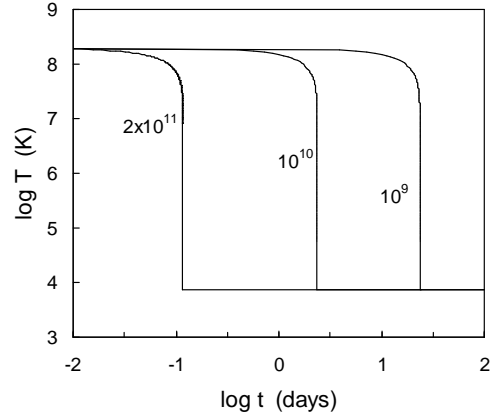}}
      \caption{Temperature of the shocked gas as a function of time for three values %%@
of the initial gas density: $10^9$, $10^{10}$ and $2\times 10^{11}$ cm$^{-3}$. In %%@
all cases the equilibrium temperature is 7500 K. 
             }
         \label{fig1}
   \end{figure}

The time dependence of the gas temperature, determined from Equation (7), is shown %%@
in figure \ref{fig1} for an initial temperature of $2\times 10^8$ K and several %%@
values of the shock gas density; the equilibrium temperature of the ionized region %%@
turned out to be in all cases $\sim 7500$ K. For lower initial temperatures the %%@
cooling time is even shorter.

Clearly, at this temperature dust grains cannot be formed or survive. However, we %%@
will show that  the two stars cannot provide enough ultraviolet photons  to ionize %%@
all the  material that accumulates behind the shock. The neutral gas will then act %%@
as a shield behind which dust can be formed. 

To quantify this statement we will consider that the matter from the $\eta$ Carinae %%@
wind accumulates behind the shock during the cooling time  (about 1 day), which %%@
corresponds, at periastron, to a thickness of the accumulated gas of about $10^{12}$ %%@
cm.
 
On the other hand, the thickness of the ionized region $\Delta r$ can be calculated, %%@
using typical O type stellar parameters, solving the equation:

\begin{equation}
\int^{\infty}_{\nu_{0}}{\frac{4\pi R^{2}_{*}F_{\nu}}{h\nu}d\nu}=\int^{r_{1}+\Delta %%@
r}_{r_{1}}{4\pi 
r^{2}n_{e}n_{\rm H}\alpha^{(2)}dr},
\end{equation}

\noindent
where $R_{*}$ is the stellar radius, $F_{\nu}$ the  flux density at the stellar %%@
surface and $\alpha^{(2)}$ the hydrogen recombination coefficient. The solution of %%@
Equation (12) gives $\Delta r \sim 6\times 10^{9}$ cm, much  smaller than the %%@
thickness of the material accumulated behind the shock front.
The  region will have a stratified structure, as shown in figure \ref{fig2}. 
At the borders (a), the stellar radiation heats and ionizes the gas. Just after this %%@
region (b) the high opacity maintains the  gas neutral. Equation (7) will still give %%@
the equilibrium temperature, but with  the photon intensity in Equation (11) given %%@
by:

   \begin{figure}
      {\includegraphics{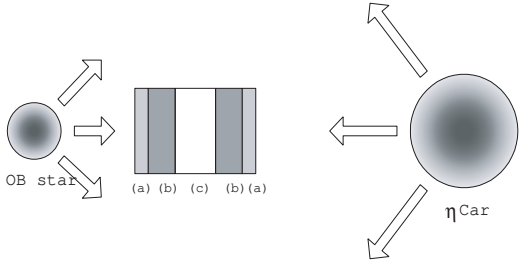}}
      \caption{Schematic view of the stratified shock region: (a) ionized region, %%@
(b) transition neutral region, (c) dust forming region}
         \label{fig2}
   \end{figure}

\begin{equation}
4\pi J_{\nu}=F_{\nu}(R_*)\frac{R_*^{2}e^{-\tau_{\nu}}}{r^{2}},
\end{equation}
where $r$ is the distance from the star and $\tau_{\nu}$  the optical depth at 
frequency $\nu$. 

The equilibrium temperature will go from $\sim 7500$ K, at the boundary with the 
ionized region, to $\sim 100$ K at the center of the shell (region "c" in fig. 2). %%@
In these conditions dust grains may be formed and also grow.

\section{Grain Formation}

Grain formation is  not a totally understood process, it envolves calculations over %%@
several elements and species, coupled to each other. To compute the kinetic %%@
equations of grain composition and evolution, the nucleation theory of moments is %%@
needed \citep*{gail84,gail87}. However, just to estimate the mean particle size and %%@
the growth timescale, the classical nucleation theory  is a good approach (Hoyle \& %%@
Wickramasinghe 1991). The dust growth and evaporation rates, described by this %%@
theory, are given by:

\begin{equation}
\left( \frac{da}{dt}\right)_{gr}=\xi\frac{1}{4\rho_{s}}\left( \frac{8kT}{\pi\mu %%@
m_{H}}\right)^{ \frac{1}{2}}\sum_{k}\frac{\rho_{k}}{A_{k}^{\frac{1}{2}}},
\end{equation}

\noindent
where $a$ is the grain radius, $\rho_{s}$  the specific density of the grain %%@
material, $A_{k}$  its atomic mass number and $\xi$ the sticking coefficient. The %%@
evaporation rate is given by:

\begin{equation}
\left( \frac{da}{dt}\right)_{ev}=-\frac{P}{\rho_{s}kT}\left( %%@
\frac{kT_{dust}m_{dust}}
{2\pi} \right)^{\frac{1}{2}},
\end{equation}

\noindent
where $P$ is the evaporation pressure.

The chemical composition of the material that will form the grains is not known. The %%@
nebula around $\eta$ Carinae is N rich and C and O poor \citep{davidson86}; infrared %%@
observations show the caracteristic 8-13 $\mu$m silicate feature \citep{robinson87}. %%@
However, {\it HST} observations detected CII and CIV lines, probably formed in the %%@
wind of the secondary star, showing a carbon-rich inner environment %%@
\citep{lamers01}.
For that reason, we will analyze the formation rates of both carbonaceous and %%@
silicate grains. 

Experimental data show that carbon dust (C$_{n}$) condensates at temperatures of %%@
about $2700$ K at the pressure $P_{sat}\sim 10^{4}$ dyn cm$^{-2}$ and that %%@
nucleation is efficient if $ P>P_{sat}$ (Draine \& Salpeter 1979; Gail, et al. 1984; %%@
Gail \& Sedlmayr 1987; Andriesse, Donn \& Viotti 1978; Hoyle \& Wickramasinghe %%@
1991). 
This condition is fully satisfied in the neutral gas region of the shell. Applying %%@
equations (14) and (15) for the post-shock 
equilibrium values we verify that equation (14) dominates over equation (15), %%@
allowing dust growth in the internal region (part "c" of figure 2). 
For large grains, evaporation becomes important and an equilibrium size for the dust  %%@
is obtained from the condition $\sum({da}/{dt}) = 0$. 
For graphite, using $\rho_{s}\simeq 2.2$ g cm$^{-3}$ (Draine \& Lee 1984), %%@
${n_{C}}/{n_{H}} \simeq 10^{-4}$  \citep{savage96} and $A_{C} = 12$, we obtained %%@
$\bar{a}\sim 0.1\; \mu{\rm m}$. The growth occurs in a time scale given by %%@
$t_{gr}\sim \bar{a}/(da/dt )_{gr}$, which is of $\sim 5$ h. It means that, after the %%@
shock is formed, the time needed for  gas cooling and further dust formation and %%@
growth to aproximately $ 10^{-5}$ cm is  of several hours. 

Considering now silicate dust formation (SiO), the saturation pressure is $P_{s} %%@
\sim 10^{2}$ dyn/cm$^2$, which gives $T_{s} \sim 1400$ K (Lefevre 1979). Considering %%@
a density $\rho_{s}\simeq 2.5$ g cm$^{-3}$ and using equations 14 and 15 we find %%@
that the grains form with a mean radius of $\bar{a}\sim 0.09\; \mu{\rm m}$ in a %%@
timescale of $\sim 10$ h. These results are similar to those obtained  considering %%@
amorphous carbon dust. 

Dust formation in the post-shock gas is also in agreement with recent observations %%@
of the binary system WR140, which shows periodic events  at each {\it periastron} %%@
passage (Monnier, Tuthill \& Danchi 2001). The same occurs in WR106, WR104 and WR137 %%@
where an increase  and consequent decay in IR flux is observed \citep*{cohen78, %%@
pitault83, cohen95, kwok97, williams01, kato02}.

\section{Shell-like efects}

The grain population formed during periastron passage must be the main source of %%@
visible and UV radiation absorption. Only recently, $HST$ observations reached %%@
enough spatial resolution to separate the stellar system from the surrounding %%@
nebula. During the 2003 low excitation event, the data presented by \citet{martin04} %%@
showed a wavelength dependent dip in the light curves for $\lambda$2200, 2500, 3300 %%@
and 5500 \AA, the corresponding absorption was 0.5, 0.38, 0.17 and 0.01 mag, %%@
compatible with the interstellar absorption law with $R \equiv A_V/E(B-V) \sim 3$ %%@
and $E(B-V) = 0.08$  mag \citep{fitz99}. 
Notice that the high absorption at  2160 \AA$\;$ in the interstellar medium is %%@
mainly due to C \citep*{mathis77}, which is  underabundant  in the $\eta$ Carinae %%@
wind, but is probably present in the wind of the companion star \citep{hillier01}. %%@
Another explanation for the larger decrease in the observed flux at the lowest %%@
wavelength, which mimics absorption by interstellar dust, is the decrease in the %%@
strength of the FeII emission lines  that also occurs during the low excitation %%@
events \citep*{gull00}, in which case a lower amount of carbonaceous grains would be %%@
needed to explain the decrease in the emission.

If the dip in the observed light curves were due to absorption by dust  with %%@
properties similar to those of the interstellar dust, the needed H column density %%@
would be \citep*{bohlin78}:
\begin{equation}
\medskip
N_H = 5.5\times 10^{21} E(B-V)\; {\rm {cm^{-2} mag^{-1}}} = 4.6\times 10^
{20} {\rm {cm^{-2}}}
\medskip
\end{equation}

As we saw in  Section 2, the H column density at the shock is about an order of %%@
magnitude  higher than what is needed to explain the UV light curve dip, implying %%@
that either the fraction of dust formed at the shock is only $10\%$ of the amount %%@
found in the  interstellar medium, or that the dust formation region  covers only a %%@
fraction of the UV emitting region.

Another consequence of the shell-like effect is the behavior of the H$\alpha$ line  %%@
and its adjacent continuum, as reported also by \citet{martin04}. While the %%@
$\lambda$6770 \AA $\;$ continuum did not show any signs of absorption during the low %%@
excitation phase, as expected from the low optical depth of the grain population at %%@
this wavelength, the H$\alpha$ emission presented a significant decrease. In the %%@
scenario presented in this paper, we interpret the variation in the H$\alpha$ light %%@
curve not as a decrease in the intrinsic emission, but as an absorption by the %%@
neutral gas in the shock front. Under these circumstances, the absorbed flux $\Delta %%@
F_{H\alpha}$, integrated across the line profile will be:

\begin{equation}
\medskip
\Delta F_{H\alpha}=\frac {h\nu _{H\alpha}}{c} B_{H\alpha}N_2  F_{H\alpha}
\medskip
\end{equation}

\noindent
where $\nu _{H\alpha}$ is the frequency of the H$\alpha$ line, $B_{H\alpha}$ the %%@
Einstein coefficient for the transition and $N_2$ the column density of H atoms in %%@
the energy level $E_2 = h\nu _2$, corresponding to the principal quantum number 2:

\begin{equation}
\medskip
N_2 = N_H e^{-h\nu _2/kT}
\medskip
\end{equation}

From the observations of \citet{martin04}, $F_{H\alpha} \simeq 5\times 10^{-9}$ erg %%@
cm$^{-2}$ s$^{-1}$ and $\bigtriangleup F_{H\alpha} \simeq 10^{-9}$ erg cm$^{-2}$ %%@
s$^{-1}$, which implies that a column density $N_2 = 10^{10}$ cm$^{-2}$ is necessary %%@
to produce the observed decrease in the H$\alpha$ flux. 
This is compatible with the inferred H column densities and temperatures of the %%@
order of 1000 K expected in the neutral region.

The interpretation of the IR light curve is complicated by the contribution of %%@
several sources \citep{whitelock94}. The amount of dust produced in our model is %%@
very small compared to the total amount of dust in the $\eta$ Carinae nebula, formed %%@
probably after the major events of mass ejection. 
Besides, in our model the dust formation process is only effective during a short %%@
time (around  periastron passage) compared to the 5.5-year orbital period. 
However, since the predicted dust temperature is high, maybe it could be related to %%@
the emission at short wavelengths observed by \citet{morris99} and \citet{hony01}, %%@
also explaining its anticorrelation  with the low excitation events seen at other %%@
wavelengths \citep{whitelock94}. 
The presence of a dip in the near infrared light curves, coincident with the minimum %%@
at other wavelengths \citep*{feast01}, can be easily explained if part of the IR %%@
emission is produced by the free-free process, in the same region that originates %%@
the optically thick millimiter spectrum.

\section{The 2-10 keV Light Curve of $\eta$ Car}

The emitted X-rays will be absorbed by matter intercepting the line of sight, %%@
regardless if it is in the form of free atoms and molecules or if it is condensed in %%@
grains.
\citet{corcoran01} were able to reproduce the X-ray light curve of the 1997 low %%@
excitation event assuming wind-wind collisions and a highly eccentric orbit. In %%@
their model, periastron  occurs when the secondary is near opposition with the Earth %%@
and the minimum in the light curve can be explained by the increase in optical depth %%@
due to the denser wind of the primary star intercepting the line of sight. However, %%@
to fit the minimum duration, they had to postulate a strong  increase in the $\eta$ %%@
Carinae mass loss rate when the distance between the two stars decreases.

In our work, we calculated the  2-10 keV X-rays emission, integrated over all the  %%@
shock surface at different orbital phases, using the expression given by  Usov %%@
(1992):

\begin{equation}
\medskip
L_{X} \simeq 2.7 \times 10^{35} \frac {\dot{M}^{1/2}_p \dot{M_s}^{3/2}}{D} \bigg( %%@
\frac {V_p^{1/2}}{ V_s^{3/2}} + 0.6\frac {V_s^{3/2}}{ V_p^{5/2}} \bigg ),
\medskip
\end{equation}
\noindent
where $\dot{M}_p$ is the mass loss rate of $\eta$ Car in $10^{-4} {M}_{\odot}\; \rm %%@
{yr}^{-1}$, $\dot{M}_s$ is the mass loss rate of the companion in $10^{-6} %%@
{M}_{\odot}\; \rm {yr}^{-1}$, $V_p$ and $V_s$ the wind speeds in units of 1000 km %%@
s$^{-1}$, and $D$ is the separation between the stars in AU. Notice the $D^{-1}$ %%@
dependence of the total flux, due to the increase in the shock surface size as the %%@
distance between the two stars increases.

However, we assumed that  periastron passage occurs when the secondary is near %%@
conjunction with the observer, as can be seen schematically in Figure 3. In fact, %%@
the actual position of the periastron relative to the observer depends on how the  %%@
observations are interpreted. \citet{damineli97} used the periodic radial velocity %%@
variations  of the Pa$\gamma$ and Pa$\delta$ emission lines to determine the orbital %%@
parameters, assuming that they were formed  in the $\eta$ Carinae wind. 
If, on the contrary, the lines  are formed in the cooling, post-shock expanding gas %%@
\citep{hill00, hill02}, the periastron will be in the opposite direction, as assumed %%@
in our work and required by the observations of \citet{davidson00}.  

In our scenario, as the secondary moves in its orbit, it leaves behind a dense and %%@
cool region, moving away from $\eta$ Carinae with velocity of about 150 km s$^{-1}$. %%@
Matter from the primary wind will be shocked to temperatures about $3\times 10^6$ K %%@
and, after isobarically cooled, it will reach densities of $2\times 10^{16}$ %%@
cm$^{-3}$. Although this material will not contribute to the observed X-ray %%@
emission, after conjunction  it will be positioned between the X-ray producing %%@
wind-wind shock  and the observer, as represented by the gray tail in Figure 3.
Notice that the absorption is due to a much larger amount of material than that %%@
which results from the wind-wind shock. The major source is the matter accumulated %%@
to the left of the secondary, after the star has passed, due to the primary wind.   

\begin{figure}
{\includegraphics{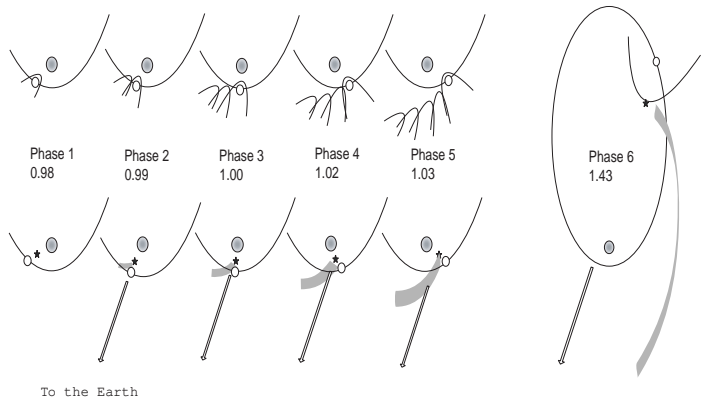}}
\caption{Evolution of the gas accumulated behind the shock, close to conjunction and %%@
periastron passage (left) and far from them (right). The upper panel represents the shocked surfaces and lower panel,  %%@
the accumulated gas (gray tail) at different points in the orbit. The numbers under %%@
the phase labels represent the true orbital phases, calculated for an eccetricity %%@
$e=0.95$ and a 5.52 year orbital period. Orbital phase one corresponds to periastron %%@
passage, which we assumed to have occurred  in December 16, 1997.}
\label{fig3}
\end{figure}

\begin{figure}
{\includegraphics{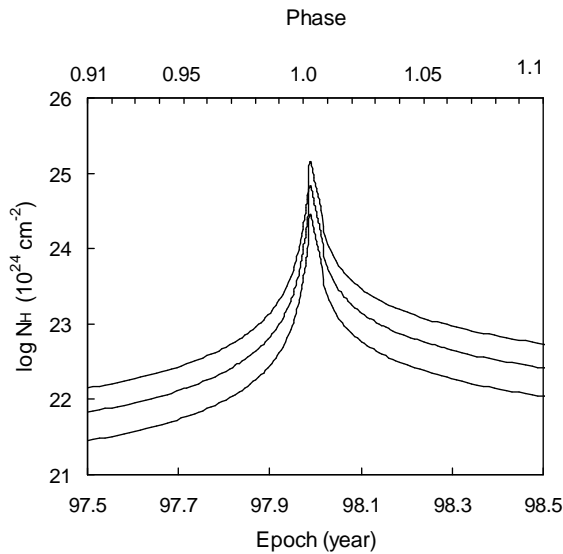}}
\caption{Neutral gas column density of the accumulated post-shocked material %%@
intercepting the line of sight at different epochs around periastron passage for the %%@
1997 low excitation event. The orbital phases shown in the upper axis were calculated %%@
with the orbital parameters described in figure 3. The three curves, in order of %%@
increasing intensity, correspond to mass loss rates for $\eta$ Carinae of 10$^{-4}$, %%@
$2\times 10^{-4}$ and $5\times 10^{-4}$ M$_{\odot}\; \rm {yr}^{-1}$.} 
\label{fig4}
\end{figure}

To calculate the amount of gas accumulated towards the line of sight at each orbital %%@
phase, we assumed that the observer is in the orbital plane of the binary system. 
Probably, this is not true (Damineli 1997), but even so, our result will not be %%@
changed appreciably if the the shell is thick.
We divided the movement of the secondary star in its orbit into time steps and %%@
computed the amount of gas intercepting the line of sight at each step, taking into %%@
account the wind contribution from $\eta$ Carinae and the fact that the shell moves %%@
away  from the star. 
The shell evolution  is schematically shown in the lower part of figure 3, phase 1 %%@
represent the starting point of our calculations; before that, the shell stays %%@
always behind the shock, and no absorption is observed. 
At phase 2, the secondary is in conjunction, and at phase 3 it reaches periastron; %%@
at this point part of the expanding shell already intercepts the line of sight and %%@
the X-rays are absorbed; at  phases 4 and 5 the column density and X-ray absorption %%@
increase. Eventually, the shell expands enough to return the absorption to its %%@
initial value. 
The resulting  column density,  as a function of epoch and phase, is shown in figure %%@
4 for three values of the $\eta$ Carinar mass loss rate. 

 The model X-ray light curves, taken into account absorption, together with the data %%@
for the 1997 low excitation event \citep{ishibashi99} are shown in figure 5. In all %%@
cases the mass loss rate of $\eta$ Carinae was taken as  $2.5 \times 10^{-4}\; \rm %%@
{M}_{\odot}\; \rm {yr}^{-1}$; in the upper panel two values for the eccentricity %%@
were used: 0.9 (light line) and 0.95 (heavy line), assuming that the periastron %%@
passage coincides with conjunction ($\Phi = 0$, where $\Phi$ is the position angle %%@
of the secondary in its orbit, measured from $\eta$ Carinae); in the lower panel, %%@
the eccentricity was taken as 0.95 and the angle between periastron and conjunction %%@
was varied from $\Phi = 0$ (heavy line) to $\Phi = 30$ (light line). We see that in %%@
all cases the model reproduces fairly well the observations.

\begin{figure}
{\includegraphics{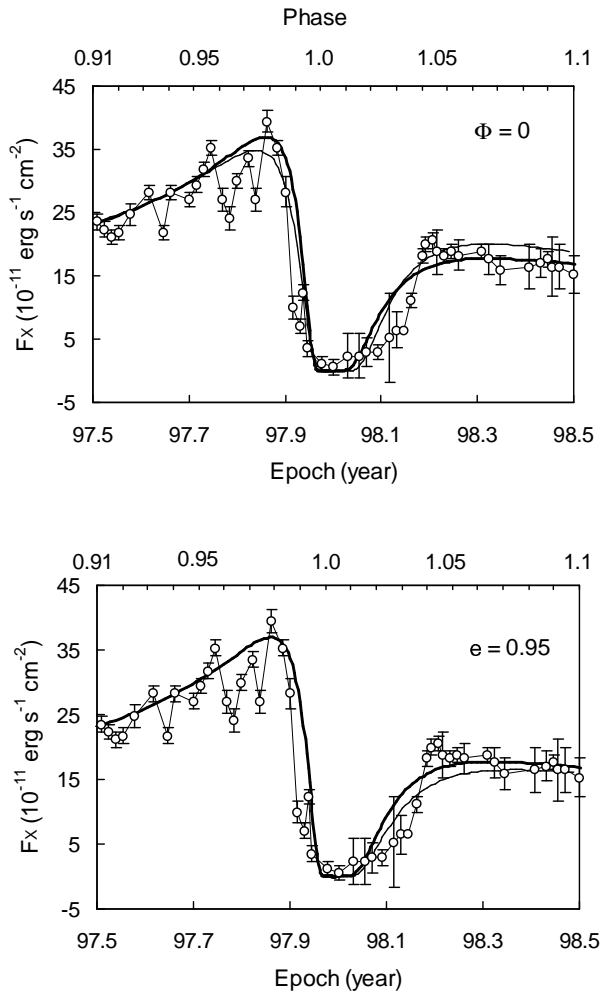}}
\caption{Model light curves for the 1997 low excitation event, superimposed to the %%@
observational data \citep{ishibashi99}, for a mass loss rate of $2.5\times 10^{-4}$ %%@
M$_{\odot}\; \rm {yr}^{-1}$. Upper panel: angle  between conjunction and periastron %%@
$\Phi$ = 0\degr; eccentricities 0.9 (light line) and 0.95 (heavy line). Lower panel: %%@
eccentricity $e = 0.95$ and angles $\Phi = 0\degr$ (heavy line) and $\Phi = 30\degr$ %%@
(light line).}
\label{fig5}
\end{figure}

\begin{figure}
{\includegraphics{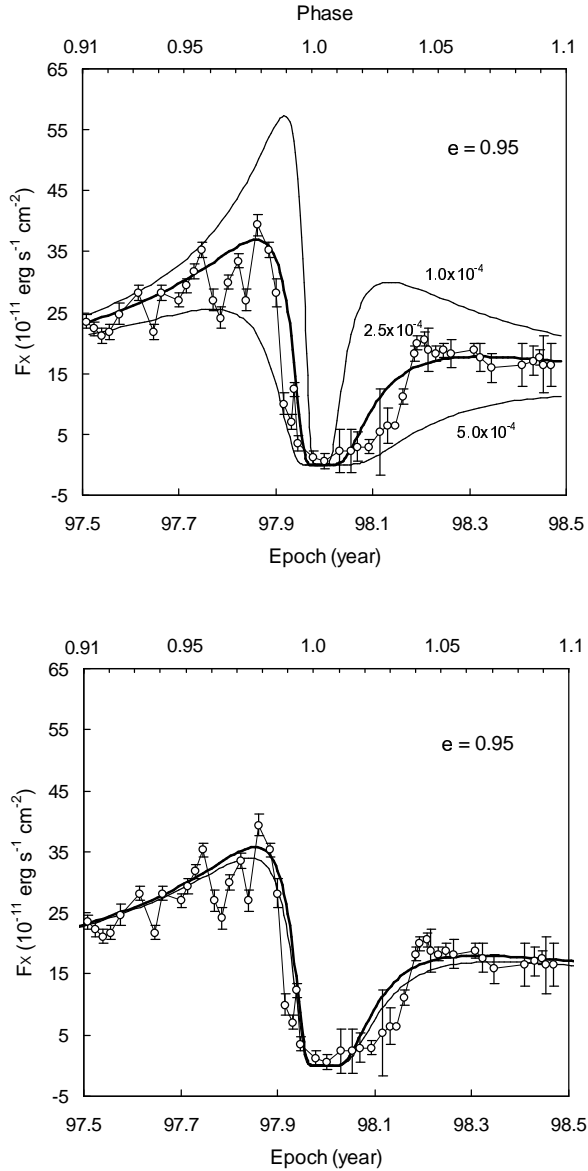}}
\caption{Model light curves for the 1997 low excitation event, superimposed to the %%@
observational data data \citep{ishibashi99} for eccentricity $e = 0.95$, angle $\Phi %%@
= 0\degr$. Upper panel: models for $v_p = 700$ km s$^{-1}$ and mass loss rates of %%@
$10^{-4}$, $2.5\times 10^{-4}$ and $5\times 10^{-4}$ M$_{\odot}\; \rm {yr}^{-1}$. %%@
Lower panel: models for mass loss rate $2.5\times 10^{-4}$ M$_{\odot}\; \rm %%@
{yr}^{-1}$ and velocities 500 km s$^{-1}$ (light line) and 700 km s$^{-1}$ (heavy %%@
line).}
\label{fig6}
\end{figure}

In the upper panel of figure 6 we compare the observations with model light curves %%@
for three values of the mass loss rate: $1.0\times 10^{-4}$, $2.5\times 10^{-4}$ and %%@
$5.0\times 10^{-4}$ M$_{\odot}\;$ yr$^{-1}$, with $v_p = 700$ km s$^{-1}$ and an %%@
orbit with $e=0.95$ and $\Phi = 0$. In the lower panel we show the model X-ray light %%@
curves for two values of the primary wind velocity : 700 km s$^{-1}$ (heavy curve) %%@
and 500 km s$^{-1}$ (light curve) for a primary mass loss rate of $2.5\times %%@
10^{-4}$ M$_{\odot}\;$ yr$^{-1}$ and orbital parameters $e=0.95$ and $\Phi = 0$. As %%@
we can see, the model light curve depends strongly on the mass loss rate and it is %%@
almost independent of the chosen  velocity. The best fitting occurs for $\dot M_p = %%@
2.5\times 10^{-4}$ M$_{\odot}\;$ yr$^{-1}$, the value obtained by \citet{pittard02} %%@
based in  numerical simulations. 

Although the calculations presented here are valid close to periastron passage, they %%@
should be able to reproduce the X-ray light curve at all phases. As the secondary %%@
moves in its orbit, the amount of material accumulated behind the shock intercepting %%@
the line of sight decreases, as shown in Figure 4. On the hand, as the star %%@
approaches opposition,  the column density of material intercepting the line of %%@
sight due to the regular $\eta$ Carinae wind increases. However, \citet{usov92} %%@
showed that the average   hydrogen column density that produces the X-ray absorption %%@
decreases as the separation between the star increases because, although absorption %%@
is large at the shock center, the larger emitting  area more than compensates this %%@
effect, as can be seen in the following equation:

\begin{equation}
\medskip
N_H\sim 10^{22}\bigg(\frac{\dot{M}_p}{10^{-5}{\rm M}_{\odot}\; {\rm %%@
yr}^{-1}}\bigg)\bigg(\frac{D}{10^{13}{\rm cm}}\bigg)^{-1}\bigg(\frac{V_p}{10^3 {\rm %%@
km \; s^{-1}}}\bigg)^{-1} 
\medskip
\end{equation}

In our scenario, at opposition $D\sim 30$ AU, resulting in a column density $N_H %%@
\sim 10^{22}$ cm$^{-2}$, too low to produce significant absorption.

\section{Conclusions}

In this work we showed that the wind-wind collision model of a binary system in a %%@
highly eccentric orbit can explain the periodic events in the 2-10 keV X-ray light %%@
curve of $\eta$ Carinae, as well as the shell-like events at optical and UV %%@
wavelengths, without any increase in  the primary mass loss rate, as postulated by %%@
\citet{corcoran01}.

As in previous works \citep{pittard98, corcoran01,pittard02}, we calculated the %%@
physical conditions in the shock region close to periastron passage, for an %%@
eccentricity $e = 0.9$,  mass loss rates of  $\dot M_p = 2.5\times 10^{-4}$ %%@
M$_{\odot}\;$ yr$^{-1}$ and $\dot M_s =  10^{-5}$ M$_{\odot}\;$ yr$^{-1}$, and wind %%@
velocities $v_p = 700$ km s$^{-1}$ and $v_s =3000$ km s$^{-1}$. We  obtained for the %%@
X-ray emitting region a density  $n_{2s}\sim 5.6 \times 10^{9}$ cm$^{-3}$ and %%@
temperature $T\sim 2\times 10^8$ K, in agreement with recent $Chandra$ observations %%@
and numerical wind collision models \citep{pittard02}.

The difference with the cited papers is that in our work we took into account that %%@
the wind from $\eta$ Carinae accumulates behind the shock, forming a shell about %%@
$10^{12}$ cm thick.
We then showed that, as the secondary moves in its orbit, the shell cools  due to %%@
free-free and line emission. Also, the number of ionizing photons from $\eta$ %%@
Carinae and its companion star are only enough to maintain the gas temperature at %%@
its equilibrium value of about 7500 K in the external parts of the shell (10$^9$ cm %%@
in depth), leaving the central part neutral and cold. We calculated that in these %%@
conditions dust grains can form and grow in timescales of a few hours. These grains %%@
would absorb the optical and UV radiation emitted by the shock material as it cools %%@
down. We used the recent $HST$ observations of the 2003 low excitation event, %%@
presented by \citet{martin04}, and showed that the decrease in the continuum flux at %%@
wavelengths of 2200, 2500, 3300 and 5500 \AA$\,$ can very well be explained in this %%@
scenario.
On the other hand, the decrease in the observed H$\alpha$ emission, without any %%@
similar effect in the adjacent continuum, was attributed by us to absorption by the %%@
neutral and cool H gas accumulated in the central region of the shell. From the %%@
amount of absorbed radiation we were able to calculate the number of hydrogen atoms %%@
in the second excited energy level ($\sim 10^{10}$ cm$^{-2}$). This is compatible %%@
with the inferred H column densities and temperatures of the order of 1000 K %%@
expected in the neutral region.

We also calculated the total amount of gas that intercepts the line of sight for %%@
each point in the orbit near periastron passage, assuming it is close to %%@
conjunction, and derived a profile similar to that found phenomenologically by %%@
\citet{corcoran01}, but without the need of an increase in mass loss rate from %%@
$\eta$ Carinae.

We constructed the expected 2-10 keV light curves for several values of the %%@
eccentricity, phase of opposition, mass loss rates  and wind velocities for $\eta$ %%@
Carinae and we were able to fit very well the shape and duration of the emission %%@
dip. We found that the mass  loss rate is the only critical parameter that affects %%@
the shape of the X-ray light curve, since for high eccentricities $\Phi$ angles lower 
than 30 degrees represent less than 1 day in the light curve timescale. The best fitting 
was obtained for $\dot M_p = %%@
2.5\times 10^{-4}$ M$_{\odot}\;$ yr$^{-1}$, which coincides with the value obtained %%@
by \citet{pittard02} in their numerical simulations. 

An interesting result of our model is that it explains the difference in the %%@
duration of the dip in the light curves at different wavelengths. At optical and UV %%@
wavelengths the dip is due to absorption by dust. Since dust growth rates are very %%@
sensitive to gas density, its formation rate, as well as the absorption it causes, %%@
will decrease very fast as the distance between the stars increases after periastron %%@
passage. X-rays, on the contrary, are absorbed by neutral gas, which will accumulate %%@
across the line of sight when the gas shell formed behind the shock expands away %%@
from the stars, and will remain there for a long time after grain formation had %%@
stopped. 

In the scenario presented here, the material accumulated in the shell will expand %%@
into the orbital plane and form a thick disk,  which can be seen at millimeter %%@
wavelengths through its free-free emission \citep{coxa95,abra99,duncan03}. The dip %%@
in the millimeter light curves can be explained by a decrease in the number of %%@
available ionizing photons, which are absorbed by the neutral shell close to the %%@
binary orbit.

Finally we must mention that what we presented here is a very simplified version of %%@
a very complicated process and any refinement in the derived orbital and physical %%@
parameters of the binary system should be obtained from detailed numerical %%@
simulations. 

\section*{Acknowledgments}

The authors would like to thank the Brazilian agencies CNPq, FAPESP and 
FINEP for partial support.

\bsp

\label{lastpage}

\end{document}